# Spin-Liquid-Like State in Pure and Mn-Doped TbInO$_3$ with Nearly Triangular Lattice

M. G. Kim[1], B. Winn[2], S. Chi[2], A. T. Savici[2], J. A. Rodriguez-Rivera[3,4], W.C. Chen[3,4], X. Xu[1], Y. Li[5,1], J. W. Kim[1], S.-W. Cheong[1], V. Kiryukhin[1]

1. Department of Physics & Astronomy, Rutgers University, Piscataway, New Jersey 08854, USA
2. Neutron Scattering Division, Oak Ridge National Laboratory, Oak Ridge, Tennessee 37831, USA
3. NIST Center for Neutron Research, National Institute of Standards and Technology, 100 Bureau Drive, Gaithersburg, Maryland 20899, USA
4. Department of Materials Science, University of Maryland, College Park, Maryland 20742, USA
5. School of Physics and Electronic Engineering, Jiangsu Normal University, Xuzhou, Jiangsu, 221116, China

**ABSTRACT**

Inelastic neutron scattering studies in single crystals of TbInO$_3$ and TbIn$_{0.95}$Mn$_{0.05}$O$_3$ with nearly-triangular antiferromagnetic lattice are reported. At low energies, a broad and apparently gapless continuum of magnetic excitations, located at the triangular lattice (TL) Brillouin zone boundary, is observed. The data are well described by the uncorrelated nearest-neighbor valence bonds model. At higher energies, a broad excitation branch dispersing from the TL zone boundary is observed. No signs of static magnetic order are found down to the temperatures two orders of magnitude smaller than the effective interaction energy. The fluctuating magnetic moment exceeds two thirds of the Tb$^{3+}$ free-ion value and is confined to the TL plane. These observations are consistent with a TL-based spin liquid state in TbInO$_3$.

# I. INTRODUCTION

Quantum spin liquids (QSL) have been continuously attracting great attention [1] since Anderson's seminal work in 1970's.[2] As most visionary ideas, QSLs have been found to be relevant to many diverse branches of physics, ranging from superconductivity[3] to quantum computing applications.[4] Known QSL candidate compounds are typically based on geometrically frustrated magnetic lattices, such as the kagome and triangular lattice.[1] QSL's can also be realized in bipartite lattices when bond-dependent anisotropic interactions are present. A prominent example is the exactly solvable Kitaev model on the honeycomb lattice,[5] which hosts Majorana fermions, and was therefore proposed as a prospective system for quantum computing. Compounds approximating this model have been recently described in the literature.[6] Unfortunately, the QSL state is notoriously difficult to detect in experiments because most of the existing probes are local, while massive entanglement and non-local excitations are thought to be the QSL's defining properties.[1] The number of strong QSL candidate compounds is very limited, and any newly discovered system is expected to attract significant interest.

The first-proposed QSL host lattice was the triangular lattice (TL).[2] After decades of activity, the properties of this lattice are still enigmatic, and therefore it remains subject of intense current research.[1] The original Anderson's idea of the QSL state in the nearest-neighbor Heisenberg spin-1/2 antiferromagnetic TL did not work, long-range magnetic order is realized instead. However, it is now widely believed that certain nearest-neighbor anisotropic couplings could induce the QSL state.[7-9] In fact, Kitaev-like QSLs have been discussed for anisotropic TLs.[7] Unfortunately, the number of strong TL QSL candidate compounds is small.[1] Examples include a few Cu-based organic compounds,[10] as well as YbMgGaO$_4$ and the related materials.[11-13] Because of the absence of the "smoking gun" experiment, the evidence of the QSL state in these systems is still not definitive. In YbMgGaO$_4$, for instance, the observed spin-liquid-like response may originate from disorder effects, and the debate on its origin is still ongoing.[14]

Herein we report a new TL-based QSL candidate system, pure and Mn-doped TbInO$_3$. It exhibits a slightly distorted triangular lattice of Tb$^{3+}$ ions, and does not exhibit any magnetic order down to 100 mK. The main technique used in this work is inelastic neutron scattering (INS). It is widely considered one of the key tools for probing the QSL state,[1,15] because it gives access to fractional magnetic excitations. Examples include spinons and the associated spinon Fermi surface in TL systems,[12] and Majorana fermions in Kitaev honeycomb lattices.[5,6] Fractional excitations are created in pairs in the INS experiment, and therefore are revealed as broad signals that are spread over a continuum of energy (the excitation continuum). Arguably, such excitations provide some of the most compelling existing evidence for the QSL, even though the data analysis may hinge on extensive theoretical modeling. In our experiments, we observe the continuum of excitations centered at the TL Brillouin zone boundary, consistent with the current understanding of TL-based QSLs. The significance of our observations may go beyond discovery of a new TL QSL candidate system, because they show that the whole series of rare earth elements, including those with even number of $f$ electrons such as Tb$^{3+}$, should be considered in the search for the QSL state.

## II. EXPERIMENT

Single crystals of TbInO$_3$ (TIO) and TbIn$_{0.95}$Mn$_{0.05}$O$_3$ (TIMO) were grown by laser floating zone (LFZ) method. High-purity Tb$_4$O$_7$, In$_2$O$_3$, and Mn$_2$O$_3$ powders were mixed with a small stoichiometric excess of In$_2$O$_3$ in a mortar, pelletized and sintered at high temperature. The sintered pellet was pulverized and made into a rod shape for the LFZ growth. The crystal was grown at the speed of 10 mm per hour in 0.8 MPa O$_2$ atmosphere. The as-grown crystal was annealed at 1400 °C for 20 hours, then cooled down to 1200 °C and then to room temperature with the rates of, respectively, 2 °C per hour and 100 °C per hour. The single crystals were inspected using an x-ray Laue diffractometer, and single crystal pieces showing sharp Laue patterns were selected for the neutron scattering experiments. Single pieces of oriented crystals of TIO and TIMO were chosen for the magnetic susceptibility measurements using a SQUID magnetometer.

The inelastic neutron scattering (INS) measurements were performed at the Hybrid Spectrometer (HYSPEC) at the Spallation Neutron Source at Oak Ridge National Laboratory, as well as at the Multi Axis Crystal Spectrometer (MACS) at NIST center for Neutron Research. Several crystals were co-aligned in the ($H$, $K$, 0) scattering plane. TIO sample of mass $m \approx 0.6$ g was studied at the HYSPEC. Incoming neutrons of energy $E_i = 7.5$ meV were used with the Fermi Chopper at the frequency $f = 180$ Hz, providing energy resolution of $\Delta E = \hbar\Delta\omega \approx 0.3$ meV at zero energy transfer $\hbar\omega = 0$. For TIMO, the HYSPEC as well as MACS were used. The same $m \approx 0.9$ g TIMO sample was studied at the both instruments. At HYSPEC, $E_i = 7.5$ meV and 3.8 meV were used, giving $\Delta E$ of 0.3 meV and 0.1 meV, respectively. At MACS, the incoming beam was doubly focused, and a cooled BeO filter was placed after the sample. Twenty double-bounce PG(002) analyzer crystals and detectors were used with a fixed final energy $E_f = 3.7$ meV, providing $\Delta E \approx 0.2$ meV resolution at $\hbar\omega = 0$. TIMO was also studied with polarized INS at MACS, using $^3$He polarizer and analyzer. The neutron polarization was perpendicular to the scattering plane and the data were corrected for the time-dependent neutron transmissions, see Ref. [16] for the details of the polarized setup. A liquid helium cryostat ($T \geq 1.7$ K), and a $^3$He dilution fridge ($T = 0.1$ K and 0.2 K) were used. For $T < 1$ K, a small magnetic field $H = 0.03$ T was applied to suppress superconductivity of the aluminum sample holder.

Indium strongly absorbs low-energy neutrons. Our composite samples had an approximately rectangular effective shape. As a result, the absorption coefficient depended of the position of the sample in the beam, as well as on the scattering angle. Therefore, anisotropic absorption corrections were required for the patterns collected in extended regions of the reciprocal space. Raw data reduction for HYSPEC and MACS was done using MANTID [17] and DAVE [18] software, respectively. Absorption corrections were calculated by modeling the transmission coefficient, first assuming the rectangular sample shape with the measured initial dimensions, and then fine tuning these dimensions so that isotropic scattering was obtained in the regions where only isotropic background was expected, such as high-energy, and elastic scattering regions. The absorption corrections were applied to all the reciprocal space maps in this paper. The corrected data match the crystallographic symmetry, showing that the correction procedure works well for both the HYSPEC and MACS data for the samples studied. In this work, none of the data are symmetrized, the actual data over the Brillouin Zone are shown. In addition, the data were corrected for the separately measured background produced by the empty sample cells.

## III. RESULTS

TbInO$_3$ crystallizes in the hexagonal structure consisting of Tb$^{3+}$ layers separated by nonmagnetic InO$_5$ layers,[19] see Fig. 1(a). Tb$^{3+}$ ions form a slightly distorted triangular lattice consisting of two inequivalent Tb sites, Tb1 and Tb2, with somewhat different *c*-axis coordinations. Tb2 forms an ideal honeycomb lattice, and Tb1 occupies the centers of the resulting hexagons, see Fig 1(b). The deviation from the ideal TL is small, the edges of the isosceles Tb triangles are only 0.6% longer than the base. The magnetic susceptibility, shown in Fig. 1(c), is highly anisotropic, indicating that the spins have strong tendency to be confined to the TL (*ab*) plane. The single-site anisotropy makes it difficult to extract the effective magnetic interaction energy from Curie-Weiss (CW) fits, and produces deviations from the CW law for the hard axis direction, *c*.[20] Nevertheless, the in-plane susceptibility is well described by the CW law below $T = 200$ K, with deviations appearing only below $T = 10$ K. The obtained $T_{CW} = -12$ K matches the typical bandwidth of the magnetic excitations obtained by INS (see below). Therefore, it gives a reliable estimate of the effective in-plane magnetic interaction, and also suggests that below $T = 10$ K magnetic moments start to correlate. The data of Fig 1(c) show no magnetic transition down to $T = 2$ K for both TIO and TIMO, and our neutron measurements indicate absence of any magnetic order down to $T = 0.2$ K (TIO), and 0.1 K (TIMO). This is two orders of magnitude smaller than the effective magnetic interaction, implying extreme frustration. Two-dimensional effects [21] play a secondary role in the nature of the ground state when the interlayer interaction is unfrustrated, as it is in TbInO$_3$. Heisenberg spins on the TL would order into the 120º structure at a temperature comparable to the interaction energy.[22] Since the next-nearest neighbor interactions between Tb$^{3+}$ ions should be weak because of the localized nature of *f* orbitals, this indicates that the magnetic frustration and lack of the magnetic order are caused by anisotropic magnetic interactions.

The microscopic nature of the magnetic state has been investigated using neutron scattering. In this paper, the Brillouin zone (BZ) boundaries and special points are marked for the undistorted Tb triangular lattice, as relevant for the results interpretation. The scattering vectors $Q = (H, K, 0)$ are quoted for the actual TIO structure having three Tb ions in the 2D unit cell shown in Fig 1(b). The data were corrected for the anisotropic neutron absorption, as described in Section II and illustrated in Fig. 2, using the TIMO sample at $T = 1.7$ K and $\hbar\omega = 0.4$ meV as an example. In TIO, a tiny nuclear Bragg peak is expected and observed at $Q = (1, 0, 0)$. Its nuclear origin is confirmed by lack of any temperature dependence, Fig. 1(d), in contrast to the strongly temperature-dependent magnetic signal. We found no other elastic magnetic peaks, sharp or broad, in the first BZ down to $T = 0.2$ K, see Figs. 1(e) and 1(f). The rings observed in Fig. 1(f) are due to a small polycrystalline TbMnO$_3$ impurity in TIMO, as confirmed by their positions, temperature dependence, and absence in the TIO samples.

Figs. 3-6 show the INS data. The main INS feature in both TIO and TIMO, best seen in Fig. 3, is a strong inelastic signal centered at the BZ boundary, with no visible dispersion up to 2 meV. This indicates a disordered but strongly fluctuating magnetic state down to $T = 0.2$ K. Figs. 3 (a-j) show constant-energy scans in the (*H, K*, 0) plane, and Figs. 4(a-c) the corresponding energy spectra along representative directions in the BZ for $T = 0.2$ K (TIO) and 1.7 K (TIMO). Figs. 3 and 4 exhibit three distinct energy regions. For $\hbar\omega < 1$ meV, a continuum of excitations is observed at the BZ boundary. For TIO, the scattering is isotropic within the experimental accuracy, see Figs. 3(a-c). In TIMO, where better counting statistics was achievable, there is a weak but noticeable bunching tendency around the M points [$Q$ = (0.5, 0.5) and equivalent], but only at the lower energies shown in Figs. 3(f), (g). We do not observe any evidence

for the excitation gap, down to our experimental accuracy of 0.1 meV, see Fig. 4 (c). Above 1 meV, there is a region of reduced intensity centered at $\hbar\omega \approx 1.3$ meV, as best seen in Figs. 4(a,b). The intensity is still centered at the BZ boundary in this region, and scattering along the boundaries of the higher BZs becomes clearly visible, see Fig. 3(i). At $\hbar\omega \geq 1.6$ meV, another scattering branch is observed. Its bottom is located at the BZ boundary and is quite isotropic, see Figs. 3(e,j). This branch is clearly dispersing away from the BZ boundary with increasing energy [Figs. 4(a,b)], in a rather isotropic manner [Fig 5]. The bandwidth of this branch, as well as of the lower-energy signal, is of the order of 1 meV (~12 K), giving an estimate for the effective interaction energy in the system. The observed features are extremely spread-out in the reciprocal space at all energies.

The temperature dependence of the inelastic scattering is depicted in Fig. 6(a), which shows powder-averaged (averaged over all $Q$ directions for each $|Q|$) signal at $\hbar\omega = 0.4$ meV. A broad maximum is observed at the $|Q|$ value corresponding to the M point of the BZ for both TIO and TIMO, suggesting a certain dominance of the corresponding scattering vector value for the low-energy regime. The signal is significantly reduced at 10 K and vanishes at 40 K. This behavior is consistent with the interaction energy scale derived from the susceptibility and INS bandwidth measurements, and suggests the magnetic origin of the INS signal. Direct evidence for such origin is provided by the polarized INS data of Fig 6(b), which shows the powder-averaged signal for $\hbar\omega = 0.8$ meV at $T = 1.7$ K. The scattering is observed only in the spin-flip channel, proving that (1) the observed INS signal is purely magnetic, and (2) the magnetic fluctuations are confined in the *ab* plane.[23] To estimate the fluctuating magnetic moment *m*, we measured the momentum-integrated dynamic susceptibility $\chi''(\omega)$ in absolute units in TIMO, using the vanadium standard for normalization, see Fig. 6(c). Using $\langle m^2 \rangle = \left(\frac{3\hbar}{\pi}\right) \int \chi''(\omega) \left(1 - e^{\frac{\hbar\omega}{kT}}\right)^{-1} d\omega = (g\mu_B)^2 J(J+1)$, and taking the $Tb^{3+}$ free ion value $g = 3/2$, one obtains a lower bound of $J = 4.0$ for the effective fluctuating moment for $\hbar\omega \leq 5$meV, which is two thirds of the full moment of $Tb^{3+}$ ion. The low-energy fluctuations therefore contain the majority of the theoretically-possible moment even if the lowest multiplet states of both Tb1 and Tb2 possess the full $Tb^{3+}$ moment. For any other scenario, practically all the moment is observed fluctuating at low temperatures. While seen in low-dimensional as well as certain frustrated magnets,[24,25] this is one of the key properties of any QSL system.

$Tb^{3+}$ has eight 4*f* electrons, and $^7F_6$ free-ion ground state with 13 degenerate levels. The local site symmetries are $C_{3v}$ and $C_3$ for Tb1 and Tb2 ions, respectively. The corresponding trigonal crystal electric field (CEF) splits the $Tb^{3+}$ levels into singlets and non-Kramers doublets. The number of these levels, their order and energies depend on the details of the local field. The observed Curie-Weiss behavior of the magnetic susceptibility with a full $Tb^{3+}$ effective moment down to $T = 10$ K excludes any singlet ground state that is separated from the next CEF level by more than a few meV. Thus, all the Tb sites are expected to be magnetic at low temperatures, either as doublets, or as singlets exhibiting singlet magnetism induced by the intersite coupling.[26] The latter situation is well known for non-Kramers rare earth ions, including $Tb^{3+}$ itself,[27] as well as $Pr^{3+}$ and $Tm^{3+}$. The data of Fig. 4(b,c) indicate a possible CEF level at 0.55 meV. The branch starting at ~1.6 meV may also be related to a CEF level, dispersing due to the comparable CEF and spin exchange interactions. Such low-lying excitations are known for $Tb^{3+}$ compounds, including the quantum spin-ice system $Tb_2Ti_2O_7$.[28-30] As already mentioned, their presence prevents any nonmagnetic Tb behavior, even if the ground state is a singlet. However, this information is insufficient to determine the nature of the $Tb^{3+}$ multiplet levels, making it difficult to construct the relevant effective microscopic

Hamiltonian. Thus, even though several existing Hamiltonian-based theoretical models based on the TL and the related honeycomb lattices predict broad scattering features at the TL BZ boundaries[9,31,32], as consistent with our data, it is too early to apply them productively. Nevertheless, theoretical calculations based on phenomenological valence bond models can still be applied, as common in both experimental and theoretical studies of QSLs.[1,33,34] We employed the nearest-neighbor uncorrelated valence bond model, and calculated the INS patterns using the equations from Ref. [33]. The result assuming the identical occupation of all the Tb-Tb dimers (the ideal TL) is shown in Fig. 3(l). To account for the possible differences in the Tb1-Tb2 and Tb2-Tb2 interactions, we have also considered the limiting cases of the Tb2-Tb2 bonds only, as well as Tb1-Tb2 bonds only, see Figs. 3(k) and (m), respectively. Interestingly, the Tb2-Tb2 bonds picture reproduces very well the lowest-energy data of Fig. 3(f), including the increased intensity at the M points. With increasing energy but still within the low-$E$ branch, the scattering is better described by the TL pattern, see Figs. 3(c,h,l). Several features typical to the Tb1-Tb2 valence bonds, such as scattering along the boundaries of the higher BZs, appear in the reduced-intensity region near 1 meV, as shown in Figs. 3(d,i,m). INS signal can only appear above the bond gap, and valence bonds with larger interaction energy are revealed at larger energy transfers. Our data are therefore consistent with the uncorrelated valence bonds model in which the Tb1-Tb2 interaction is stronger than the Tb2-Tb2 interaction.

Phenomenologically, our INS data for $\hbar\omega < 1.5$ meV look very similar to those obtained for the TL QSL candidate YbMgGaO$_4$, in which the Yb$^{3+}$ ions are Kramers doublets, and both the spinon Fermi surface[12] and the valence bond scenario have been considered.[33] Resonating valence bond models can realize the QSL, but the bonds may also be static, similar to the correlations found in frozen frustrated states.[34]-[37] It has been argued that the INS response in YbMgGaO$_4$ and YbZnGaO$_4$ only mimics the QSL,[14,25,34] and the observed broad features are induced by the unavoidable (Mg,Zn)/Ga disorder. In contrast, there are no known sources of disorder in the pure TbInO$_3$. The observed combination of large magnetic frustration, lack of any static order, essentially fully fluctuating magnetic moment, and the low-energy broad and dispersionless continuum of excitations strongly indicate a non-trivial fluctuating magnetic ground state in TbInO$_3$.

The two distinct Tb sites in TbInO$_3$ make the system more complex than the ideal TL, but they also expand the range of the possible properties. The Tb1/Tb2 lattice of TbInO$_3$ is of the type known as stuffed honeycomb lattice.[38] For isotropic spin-half lattices of this type, various ordered states, as well as QSLs, have been predicted.[38] Further theoretical work on anisotropic interactions in this model should be relevant to TbInO$_3$, and is therefore highly desirable. An important limit of the stuffed lattice is the simple honeycomb lattice, which can support the Kitaev QSL with its exotic dynamic properties, such as Majorana fermions. As described above, valence bonds modeling instead supports the scenario not too remote from the TL limit, with the weaker Tb2-Tb2 interaction that moves the system away from the honeycomb limit. Another feature consistent with the TL-like scenario is the enhanced scattering near the M points in the BZ observed at low energies in TIMO. It indicates short-range dynamic correlations at the wave vector corresponding to the stripy magnetic order, which is well known in the TLs frustrated either by anisotropic interactions (as probably relevant to our case),[7-9] and in the isotropic systems with next-neighbor interactions.[32] We are not aware of any ordered state on the honeycomb lattice (including stripy, zig-zag, *etc*) with peaks at the M points that also do not have strong scattering at smaller wave vectors, which is not observed in our experiments. Thus, while the actual magnetic state of TbInO$_3$ may interpolate between the TL, honeycomb, or some other scenarios, our data indicate that the low-energy dynamic response is well approximated by the physics of the triangular lattice.

## IV. CONCLUSIONS

The magnetic subsystem in pure and Mn-doped TbInO$_3$ does not show any evidence of the static order down to the temperatures two orders of magnitude smaller than the typical interaction energy. Instead, the large fraction of the magnetic moment (at least two thirds of it for $\hbar\omega \leq$ 5meV and $T$ = 1.7 K in TIMO) remains fluctuating in the plane of the triangular lattice. Our INS data show broad excitation continuum centered at the TL BZ boundary. They are consistent with the uncorrelated nearest-neighbor valence bonds model on the TL with somewhat reduced interaction for the honeycomb (Tb2) sublattice. Thus, several key features associated with the QSL state are present in TbInO$_3$. Importantly, the pure compound does not have any known sources of disorder, in contrast to the well-known TL compound YbMgGaO$_4$, and therefore the observed behavior is probably intrinsic. Given the paucity of the QSL candidate compounds, we expect TbInO$_3$ to attract significant interest. Importantly, the observed fluctuating magnetic state is realized in the non-Kramers Tb$^{3+}$ system with eight 4$f$ electrons. Our measurements therefore confirm that candidate QSL compounds could be found in the rare earth compounds with both odd and even number of the $f$ electrons.

*Note added*. On manuscript submission, we became aware of the article by L. Clark et al. [39] reporting spin-liquid behavior in polycrystalline samples of TbInO$_3$. Our results are in agreement with one major conclusion of Ref. [39], namely the spin-liquid-like behavior of TbInO$_3$ at low temperatures. Our INS measurements were done on single crystals, and therefore provide more detailed information than those reported in [39]. In particular, our data do not support the low-temperature emergence of the honeycomb lattice (weak Tb1-Tb2 bond limit), which is another major conclusion of Ref. [39].

### Acknowledgements

Work by the Rutgers group, including sample growth and neutron scattering, was supported by the U.S. Department of Energy (DOE) under Grant No. DOE: DE-FG02-07ER46382. A portion of this research used resources at the High Flux Isotope Reactor and the Spallation Neutron Source, a DOE Office of Science User Facility operated by the Oak Ridge National Laboratory. Access to MACS was provided by the Center for High Resolution Neutron Scattering, a partnership between the National Institute of Standards and Technology and the National Science Foundation under Agreement No. DMR-1508249. We thank C. D. Batista for useful discussions.

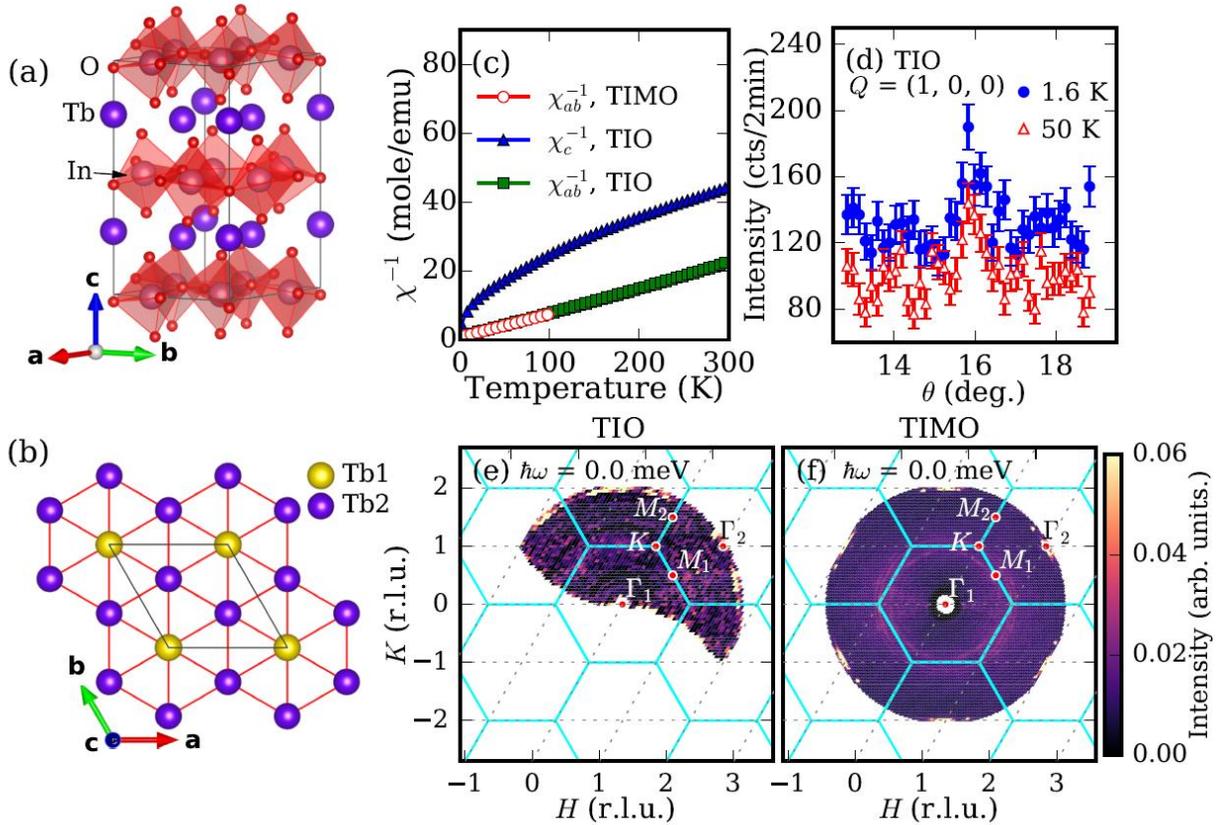

**Figure 1.** (a) Crystallographic structure of TbInO$_3$. (b) Tb1 and Tb2 sites in each Tb layer. Gray lines show the crystallographic unit cell. (c) Anisotropic inverse magnetic susceptibility for TbInO$_3$, and TbIn$_{0.95}$Mn$_{0.05}$. (d) Elastic sample rocking scans through the (1,0,0) nuclear peak position in TbInO$_3$ for $T$ = 1.6 K, and 50 K. Error bars represent one standard deviation. (e),(f) Elastic neutron scattering patterns collected in TbInO$_3$ for $T$ = 0.2 K (d), and in TbIn$_{0.95}$Mn$_{0.05}$O$_3$ for $T$ = 1.7 K (e). Solid lines delineate the BZ boundaries of the TL, the special BZ points are marked.

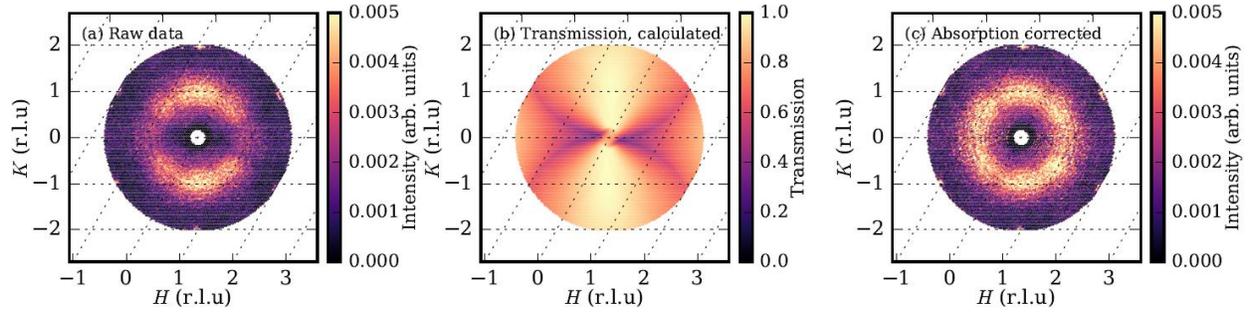

**Figure 2.** Scattering pattern collected in TbIn$_{0.95}$Mn$_{0.05}$O$_3$ at the HYSPEC instrument for $\hbar\omega = 0.4$ meV and $T = 1.7$ K in a liquid helium cryostat. (a) Raw data. (b) Sample transmission, calculated as described in the text. (c) Absorption-corrected data.

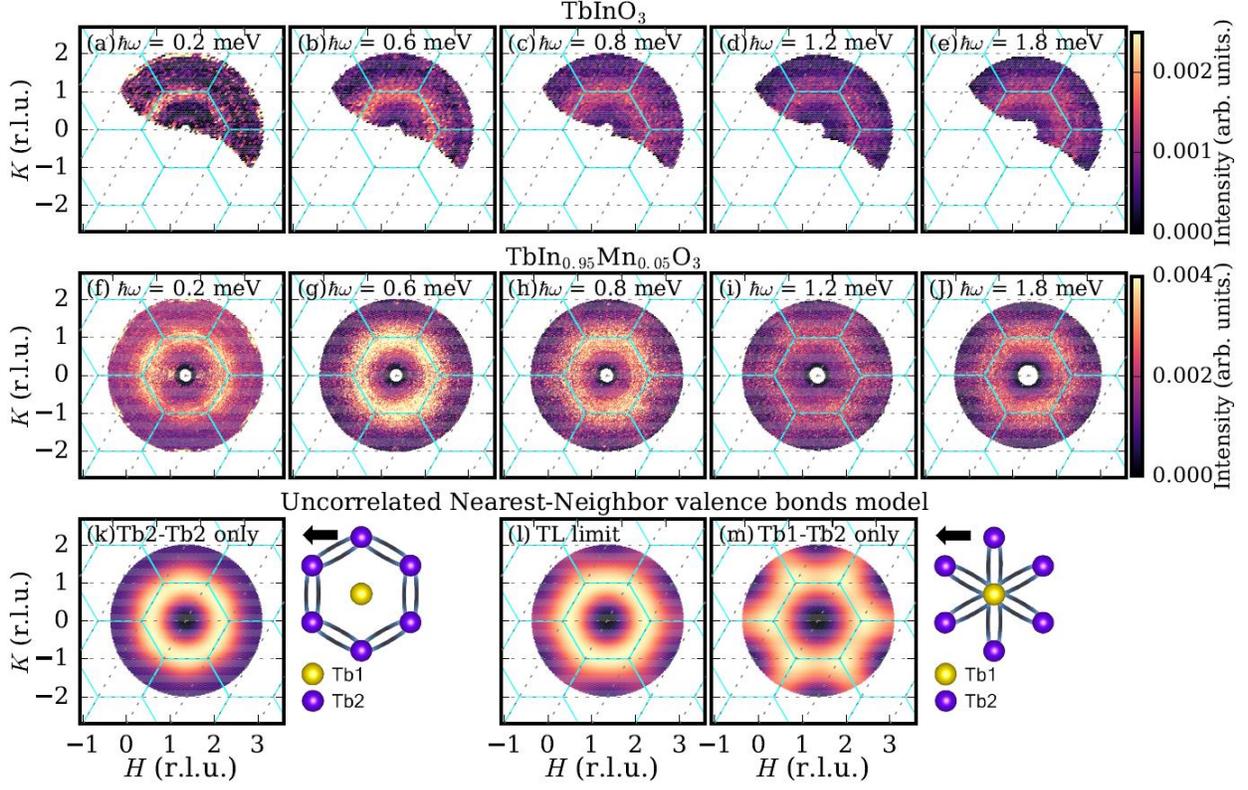

**Figure 3.** Inelastic neutron scattering patterns in TbInO$_3$ at $T$ = 0.2 K (a)-(e), and TbIn$_{0.95}$Mn$_{0.05}$O$_3$ at $T$ = 1.7 K (f)-(j) for different energy transfers $\hbar\omega$. The data were taken at the HYSPEC using 0.3 meV resolution and integrating over 0.2 meV window. Intensity units are arbitrary. Solid lines delineate the BZ boundaries of the TL. Nearest-neighbor valence bonds calculations for the Tb2-Tb2 bonds only, for the equally-occupied bonds (the TL limit), and for the Tb1-Tb2 bonds only are shown in (k), (l), and (m), respectively. The cartoons sketch the positions of the dimers for the cases (k) and (m).

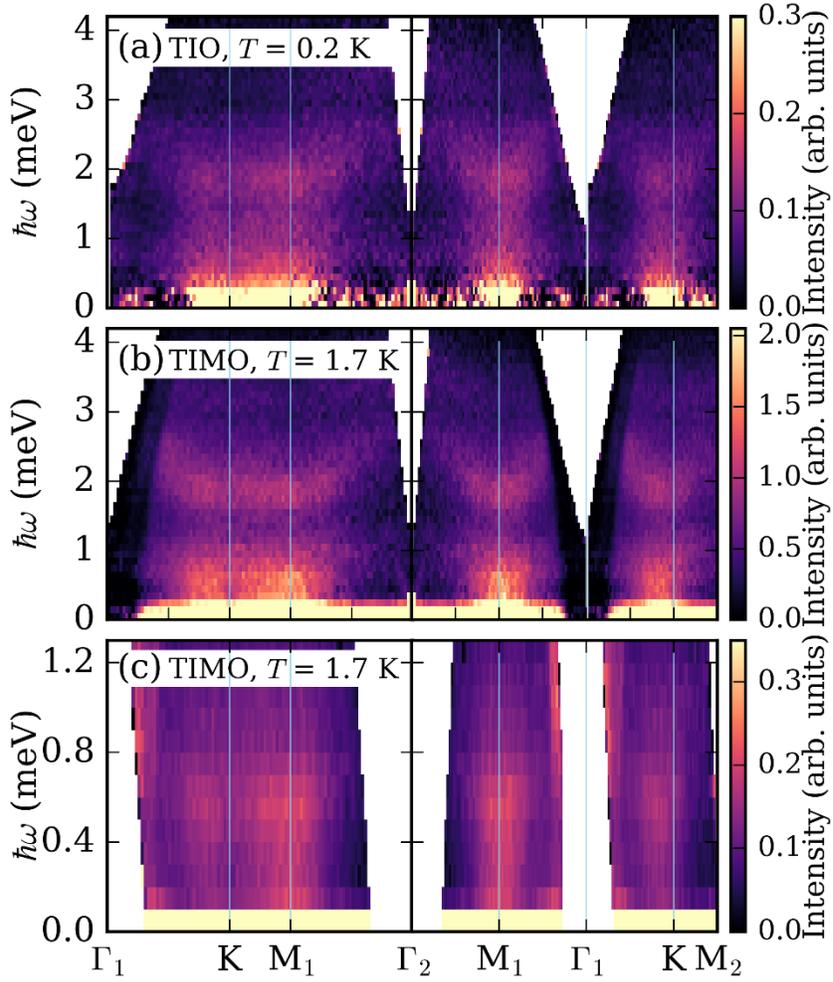

**Figure 4.** INS spectra for several special directions in the TL BZ, taken at HYSPEC for TbInO$_3$ at $T = 0.2$ K (a), and TbIn$_{0.95}$Mn$_{0.05}$O$_3$ for $T = 1.7$ K (b), (c). The energy resolution is 0.3 meV in (a), (b), and 0.1 meV in (c).

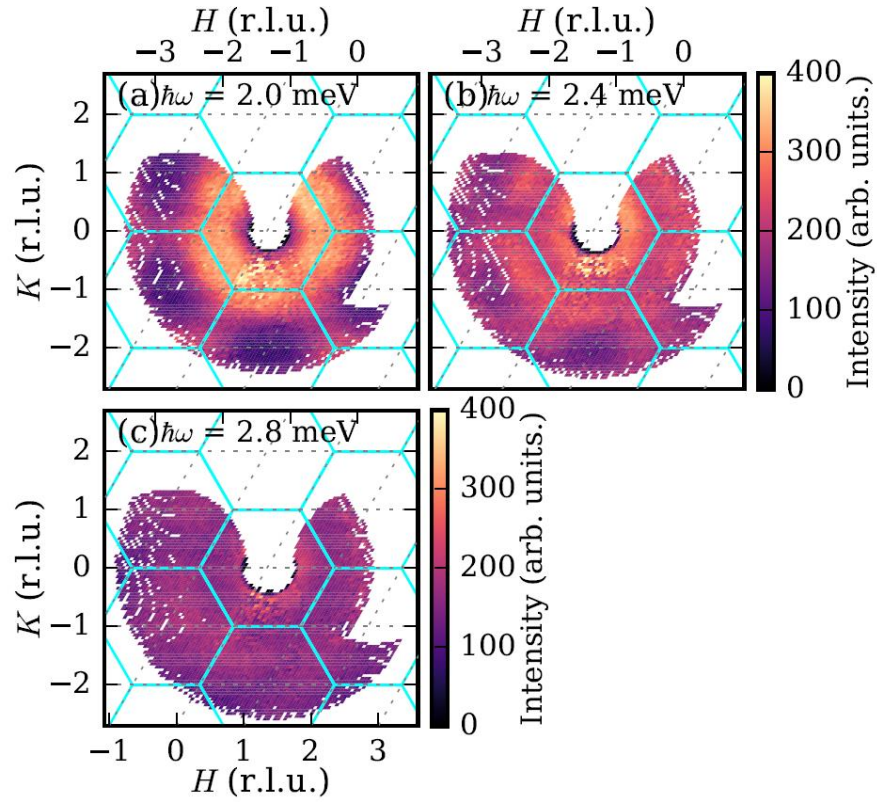

**Figure 5**. INS patterns collected in TbIn$_{0.95}$Mn$_{0.05}$O$_3$ at $T = 1.7$ K for several higher-energy energy transfers $\hbar\omega$ at the MACS instrument. Solid lines delineate the BZ boundaries of the triangular lattice. The signal is clearly dispersing away from the BZ boundary with increasing energy.

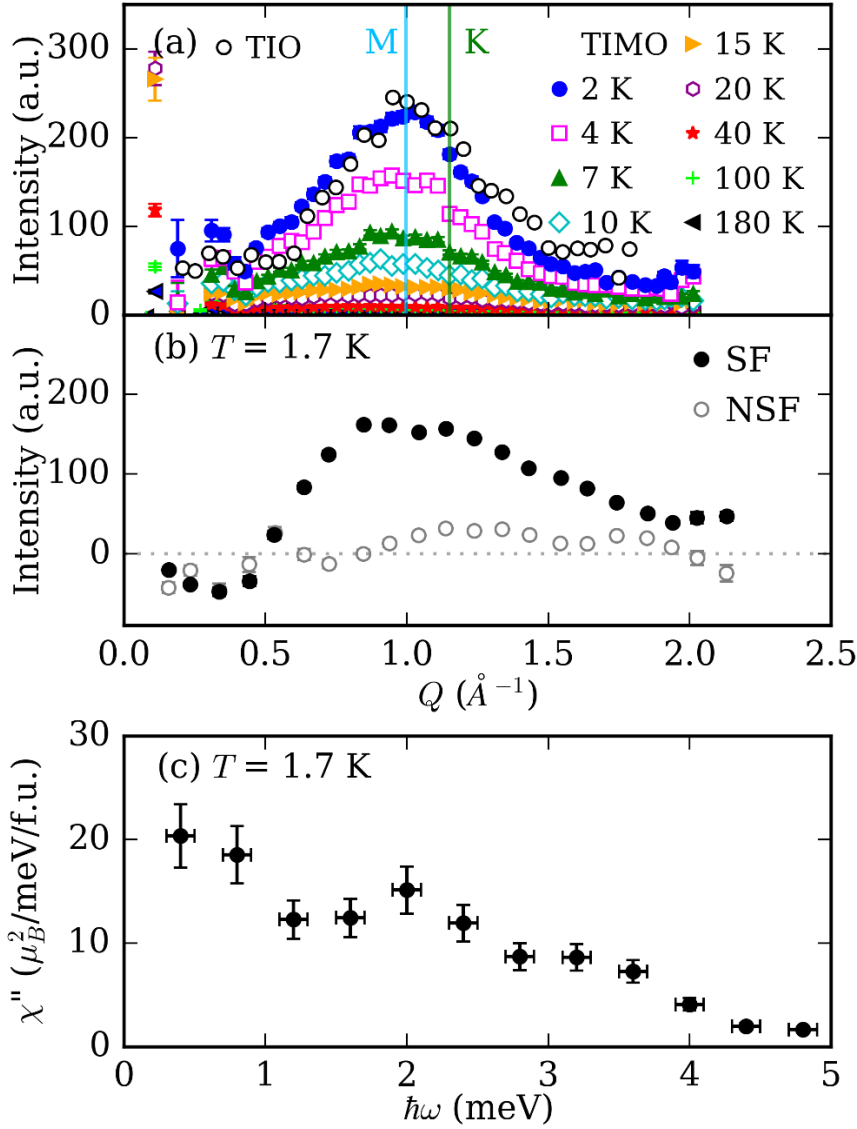

**Figure 6.** (a) Powder-averaged neutron scattering intensity for $\hbar\omega = 0.4$ meV at $T = 0.2$ K for TbInO$_3$, and at several temperatures for TbIn$_{0.95}$Mn$_{0.05}$O$_3$. Vertical lines show the positions of the M and K points in the TL BZ. (b) Powder-averaged *polarized* neutron scattering intensity for $\hbar\omega = 0.8$ meV and $T = 1.7$ K in TbIn$_{0.95}$Mn$_{0.05}$O$_3$. Solid and open symbols show spin-flip (SF) and non-spin-flip (NSF) channels, respectively. (c) Momentum-integrated dynamic susceptibility in the absolute units for $T = 1.7$ K in TbIn$_{0.95}$Mn$_{0.05}$O$_3$. Error bars represent one standard deviation.